\documentclass[aps,prl,twocolumn,showpacs,groupedaddress]{revtex4-1}

\usepackage{graphics}

\usepackage{epsfig}
\usepackage{amsmath}
\usepackage{amssymb}
\usepackage{amsfonts}
\usepackage{ dsfont }

\usepackage[debug]{mciteplus}

\usepackage{color}
\usepackage{ulem}

\newcommand{\be}{\begin{equation}}
\newcommand{\ee}{\end{equation}}

\newcommand{\ket}[1]{\mbox{$|#1\rangle$}}

\begin{document}

\hsize\textwidth\columnwidth\hsize\csname@twocolumnfalse\endcsname

\title{Interfacing spin qubits in quantum dots and donors - hot, dense and coherent}

\author{ L. M. K. Vandersypen$^{1,2}$, H. Bluhm$^3$, J. S. Clarke$^2$, A. S. Dzurak$^4$, R. Ishihara$^1$, A. Morello$^4$, D. J. Reilly$^5$, L. R. Schreiber$^3$, M. Veldhorst$^1$\\}

\affiliation{$^1$ QuTech and Kavli Institute of Nanoscience, TU Delft, Lorentzweg 1, 2628CJ Delft, the Netherlands\\
$^2$ Components Research, Intel Corporation, 2501 NW 29th Ave, Hillsboro, OR 97124, USA\\
$^3$ JARA-Institute for Quantum Information, RWTH Aachen University, D-52074 Aachen, Germany\\
$^4$ Centre for Quantum Computation and Communication Technology, School of Electrical Engineering and Telecommunications, The University of New South Wales, Sydney NSW 2052, Australia\\
$^5$ ARC Centre of Excellence for Engineered Quantum Systems, School of Physics, The University of Sydney, Sydney, NSW 2006, Australia}

\date{\today}

\vskip1.5truecm

\begin{abstract}
Semiconductor spins  are one of the few qubit realizations that remain a serious candidate for the implementation of large-scale quantum circuits. Excellent scalability is often argued for spin qubits defined by lithography and controlled via electrical signals, based on the success of conventional semiconductor integrated circuits. However, the wiring and interconnect requirements for quantum circuits are completely different from those for classical circuits, as individual DC, pulsed and in some cases microwave control signals need to be routed from external sources to  every qubit. This is further complicated by the requirement that these spin qubits currently operate at temperatures below 100 mK. Here we review several strategies that are considered to address this crucial challenge in scaling quantum circuits based on electron spin qubits. Key assets of spin qubits include the potential to operate at 1 to 4 K, the high density of quantum dots or donors combined with possibilities to space them apart as needed, the extremely long spin coherence times, and the rich options for integration with classical electronics based on the same technology.
\end{abstract}

\maketitle

The quantum devices in which quantum bits are stored and processed will form the lowest layer of a complex multi-layer system \cite{Metodi2005,Jones2012,Fu2016}. The system also includes classical electronics to measure and control the qubits, and a conventional computer to control and program these electronics. Increasingly, some of the important challenges involved in these intermediate layers and how they interact have become clear, and there is a strong need for forming a picture of how these challenges can be addressed. 

Focusing on the interface between the two lowest layers of a quantum computer, each of the quantum bits must receive a long sequence of externally generated control signals that translate to the steps in the computation. Furthermore, given the fragile nature of quantum states, large numbers of quantum bits must be read out periodically to check whether errors occurred along the way,  and to correct them \cite{Terhal2015}. Such error correction is possible provided the probability of error per operation is below the accuracy threshold, which is around $1 \%$ for the so-called surface code, a scheme which can be operated on 2D qubit arrays with nearest-neighbour couplings \cite{Raussendorf2007,Fowler2012}. The read-out data must be processed rapidly and fed back to the qubits in the form of control signals. Since each qubit must separately interface with the outside world, the classical control system must scale along with the number of qubits, and so must the interface between qubits and classical control. 

The estimated number of physical qubits required for solving relevant problems in quantum chemistry or code breaking is in the $10^6-10^8$ range, using currently known quantum algorithms and quantum error correction methods \cite{Wecker2014, VanMeter2013}. For comparison, state-of-the-art processors contain more than one billion transistors \cite{Inteltransistorcount}. Furthermore, the structure of these transistors bears a lot of resemblance with that of promising semiconductor based qubits \cite{Hanson2007, Zwanenburg2013}. However, an important difference is that conventional processor chips have only $\approx 10^3$ input-output connections (IO's), for instance Intel's land grid array (LGA) 2011 socket has 2011 pins that contact the backside of the processor \cite{LGA2011}. This brings the transistor-to-IO ratio over $10^6$. This scaling of the number of pins with the number of devices is empirically described by Rent's rule \cite{Lanzerotti2005}. In the absence of multiplexing or on-chip control logic, the limit for the qubit count is probably similar to the pin-limit of the package, which is currently around $10^3$ \cite{LGA2011}.

Therefore, the notion that semiconductor quantum bits that are manufactured by CMOS-compatible technology are easily scalable, is too simplistic. While many qubit architectures and strategies for scaling have been proposed \cite{Loss1998, Kane1998, Childress2004, Taylor2005a, Burkard2006, Hollenberg2006, Friesen2007, Trauzettel2007, Hermelin2011, McNeil2011, Hu2012, Shulman2012, Trifunovic2012, Trifunovic2013, Leijnse2013, Hill2015, Hassler2015, Viennot2015, Stano2015, Srinivasa2015, Schuetz2015, Tosi2016, Viennot2015, Baart2016a, Baart2016b, Pica2016, Veldhorst2016, Gorman2016, Petersson2012}, a completely worked out pathway to create qubit systems that can be expanded to a large-scale quantum processor yet has to be defined and a key step is the design of a scalable classical-quantum interface.

Here we focus on the quantum-classical interface requirements and possible solutions for qubits encoded in electron spins in semiconductor quantum dots and donors \cite{Hanson2007, Zwanenburg2013}. We thereby consider specifically quantum dots that are probed and controlled using electrical signals, referring to \cite{Jones2012} for a discussion of optically addressed quantum dots. Electrically controlled quantum dots and donors are two promising qubit realisations that have much in common both conceptually and in terms of qubit specifications and hardware requirements. There is significant scope to make these realisations compatible with industrial complementary-metal-oxide-semiconductor (CMOS) technology, which is optimized for high-yield, reproducibility and cleanliness. Indeed, there is a lot of effort in this direction and qubits that are partly fabricated with industrial technology have already been realized \cite{Maurand2016}. 

We begin with a brief summary of electron spin qubits in quantum dots and donors, then derive the control signal requirements and challenges, and next present possible solutions to overcome these challenges. These focus on dense 2D tunnel coupled spin qubit arrays, sparse arrays with coherent links between them, and on the possibility of operating spin qubits at 1 or 4 K, allowing for more complex electronics to be integrated with the qubits.

\begin{figure*}
\includegraphics[width=0.9\textwidth]{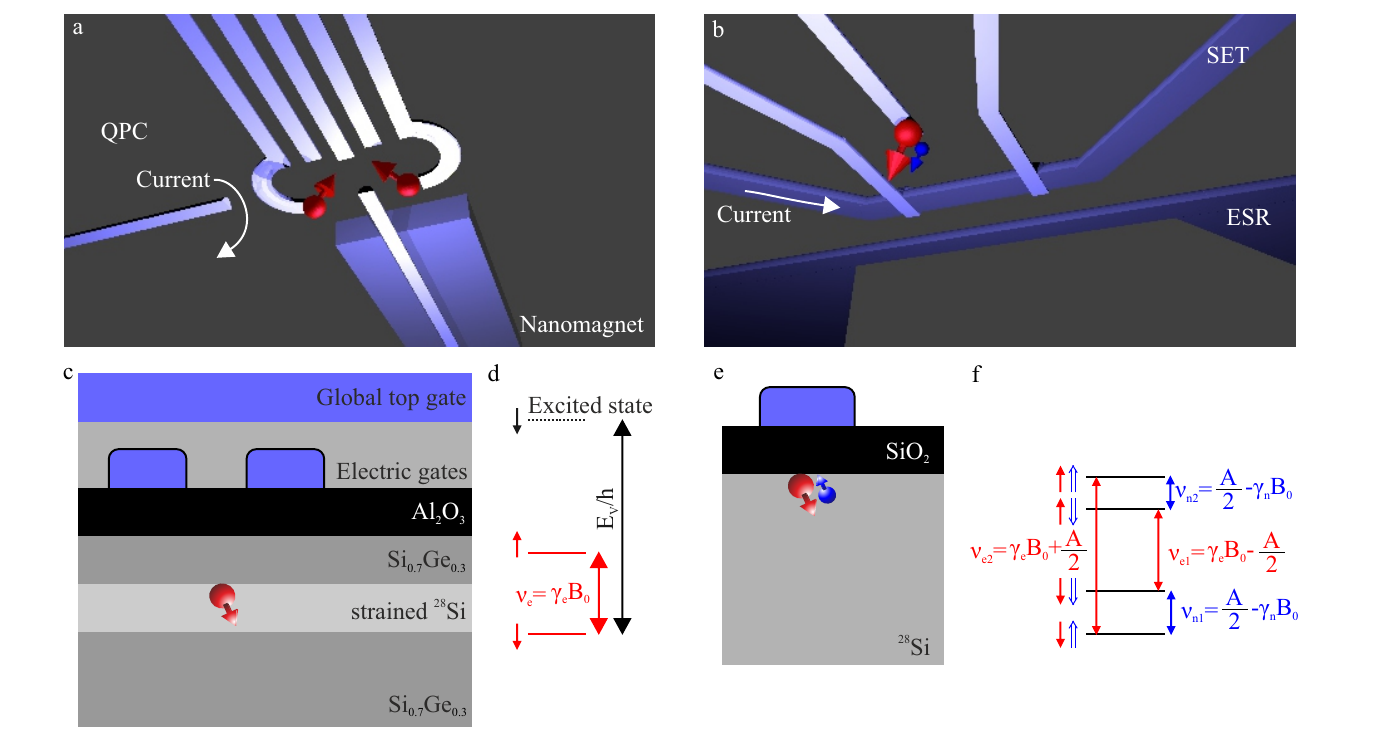}%
\caption{Schematic diagram of typical electrically measured spin qubit devices. (a) A double quantum dot device defined in a Si/SiGe quantum well. Quantum dots can be defined either in accumulation mode with a global top gate as depicted in panel (c), or in depletion mode using a doping layer. (b) Donor qubit system in depletion mode and fabricated by silicon metal-oxide-semiconductor technology (material stack in e). The spin states of a single electron are split in a magnetic field and qubit operation is obtained via an ac magnetic field that matches the associated resonance frequency $\nu_e$ as represented in (d) for dots and (f) for donors. An ac magnetic field can be realized directly via a strip-line (b). Alternatively, the motion of a quantum dot due to an ac electric field created by a nearby gate results in an effective magnetic field due to the field gradient of a nearby nanomagnet (a). The donor system forms an effective two-qubit device due to the presence of a nuclear spin, that is coupled to the electron through the hyperfine interaction with strength $A$. The gyromagnetic ratio $\gamma$ of both the quantum dot and donor system are affected by the electric field from the nearby electrostatic gates and nearby charged defects, which causes a non-uniformity between the qubits, but can also be exploited for addressability. For high-fidelity operation it is important that the qubit states are well isolated from excited states. Particularly in silicon quantum dots, a low-energy excited state can appear due to valley degeneracy, which can be lifted in energy via a large vertical electric field \cite{Yang2013}. The quantum-point-contact (QPC) or single-electron-transistor (SET) is used to probe the number of charges on the dots. They could potentially be avoided via gate-based dispersive read-out \cite{Colles2013}. 
\label{fig:dot-donor}}
\end{figure*}

\section{Electron spin qubits in quantum dots or donors} 
\label{sec:spinqubits}

\begin{figure*}
\includegraphics[width=0.9\textwidth]{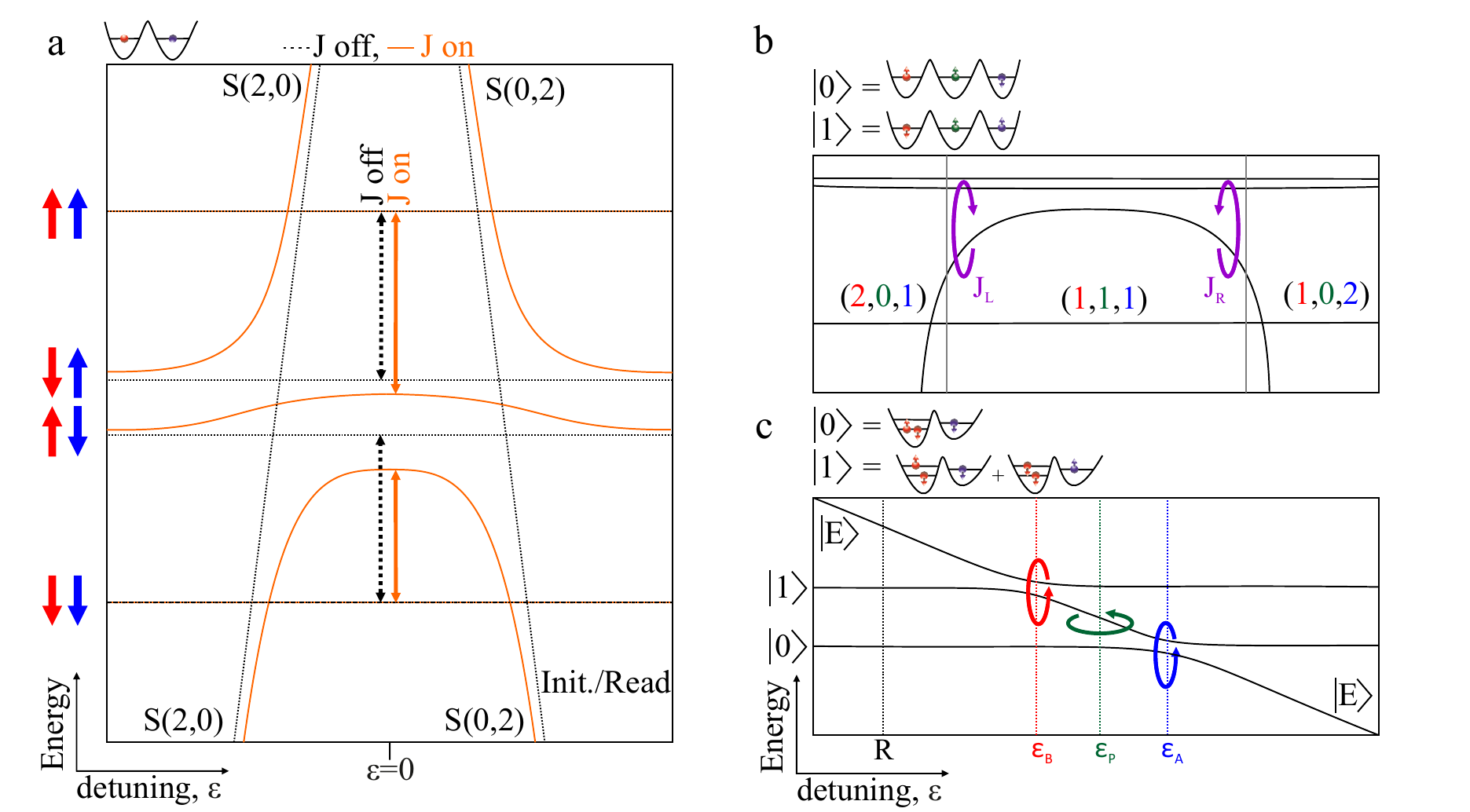}
\caption{Energy level diagram of spin states in quantum dots. (a) Low-energy spectrum of two uncoupled spins (black dotted line) and coupled spins (orange solid line) in two quantum dots as a function of the detuning energy $\epsilon$, the relative energy difference between the left and right dot levels, which is controlled by the corresponding dot gate voltages. The exchange interaction provided by the charge states with double occupancies ($S(2,0)$ and $S(0,2)$) can be used for two-qubit operations between single spin qubits as the exchange interaction $J$ modifies the qubit resonance frequencies. While in the uncoupled situation the transition $\ket{\downarrow\downarrow}$ to $\ket{\uparrow\downarrow}$ has the same energy as the $\ket{\downarrow\uparrow}$ to $\ket{\uparrow\uparrow}$ transition, these become different when exchange is on, allowing to drive rotations of one spin conditional on the state of the other \cite{Veldhorst2015}. Alternatively, when briefly turning on the exchange, the two spin states will exchange over time, which also constitutes a two-qubit gate. While many experimental works exploits the detuning to control the exchange amplitude, directly controlling the tunnel coupling allows to operate the system at the so-called symmetry point, where the exchange energy is less sensitive to charge noise, dramatically improving the gate fidelity \cite{Reed2016,Martins2016}. The joint state of two coupled spins, for instance the spin singlet and one of the triplet states, can also be used as a single qubit \cite{Petta2005}. The advantage of such a qubit is that one qubit axis is electrically controlled and two qubits can be coupled capacitively \cite{Shulman2012}. For universal control, a magnetic field gradient is required, for instance induced by a nearby nanomagnet. All electrical control is possible using more advanced combinations of spins, for example (b) the so-called exchange-only qubit and (c) hybrid qubit. (b) The encoding in the exchange-only qubit is based on three spins in three adjacent quantum dots and control is provided via the exchange between the outer quantum dots and the central dot, $J_L$ and $J_R$ \cite{Laird2010, Medford2013, Eng2015}. (c) The hybrid qubit is based on three spins as well, but requires only two quantum dots \cite{Kim2014}. Universal qubit control makes use of the anti-crossings between the lowest three energy states to induce rotations about different axes. While these qubit representations are clearly more involved compared to the single-spin qubit, their operation may offer advantages for scaling towards large arrays where not the number of dots per qubit but the number and type of control lines per dot will likely form the largest challenge.
\label{fig:encodings}}
\end{figure*}

We first briefly introduce electron spin qubits in electrically detected quantum dots and donors as a starting point for discussing the control and interfacing requirements (for more extensive reviews, see \cite{Hanson2007, Zwanenburg2013}). 

A schematic of a prototypical quantum dot device is shown in Fig.~\ref{fig:dot-donor}(a). A combination of bandgap offsets and electrostatic gates are used to confine one or more free electrons (or holes \cite{Maurand2016, Li2013, Spruijtenburg2013}; for brevity we will refer to electrons throughout the text) in a small space in a semiconductor, typically a few tens of nm in diameter. For qubit experiments, the gate voltages are usually tuned so the quantum dots contain exactly one electron each, although for certain initialization and read-out protocols, an electron is pushed off a dot or onto a neighbouring dot. Fig.~\ref{fig:dot-donor}(b) shows a schematic of a donor-based device. Donor atoms such as phosphorous in silicon have one excess electron compared to the atoms in the surrounding lattice, and at low temperatures this electron is bound to the donor atom (acceptors with one excess hole can be used as well; we will just refer to donors for brevity). With a gate voltage, this electron can be pushed off the donor or a second electron can be bound to the donor, provided the required electric fields are below values that result in population of the silicon conduction band (or valence band in case of acceptors). In both cases, an additional gate can be placed in between or close to adjacent sites in order to control the tunneling of electrons between the sites via a gate voltage, a crucial ingredient of most electron spin qubit proposals \cite{Loss1998,Kane1998}. Qubit experiments with such systems have been performed so far with the sample attached to the mixing chamber of a dilution refrigerator, at operating temperatures of 10-100 mK.

The canonical encoding of a qubit in these systems is in the spin split levels, $\ket{\uparrow}$ and $\ket{\downarrow}$, of the electron on each site, in the presence of a static magnetic field \cite{Loss1998,Kane1998}. However, alternative encodings have been proposed theoretically and explored experimentally, whereby specific collective spin states of two or three electrons in two or three quantum dots are used to represent $\ket{0}$ and $\ket{1}$, see Fig.~\ref{fig:encodings} \cite{DiVincenzo2000,Levy2002, Laird2010, Koh2012, Kim2014}. For each of these encodings, DC voltages may be used to fine tune qubit transition frequencies. This is immediate for the encodings based on two or three electron spins, where qubit splittings are directly set by gate voltages. However, for single-spin qubits, the spin splitting is also typically sensitive to electric fields \cite{Kawakami2014, Veldhorst2014, Laucht2015}.

Regardless of the chosen qubit encoding, one generally requires the ability to individually rotate every qubit about two different axes in the corresponding qubit Bloch sphere, and to entangle neighbouring qubits with each other; see Fig.~\ref{fig:encodings} for rotation axes of different qubit encodings. Together, this forms a universal set of quantum gates, which can be used to perform arbitrary logic \cite{Nielsen2000}. Both single-qubit and two-qubit gates can be accomplished in one of two modes: (1) fast gate voltage pulses that rapidly switch the Hamiltonian so that the qubit(s) will start evolving around a new axis (in Hilbert space) or (2) radio-frequency or microwave electric or magnetic fields resonant with the energy difference between specific single- or two-qubit states. Gate durations vary from sub-ns to microsecond timescales \cite{Hanson2007, Zwanenburg2013}.

Spin states are hard to detect directly, but can be converted to charge via a sequence whereby a charge movement between dots or from a dot to a nearby electron reservoir is made to be spin state dependent, via ``spin-to-charge conversion" \cite{Ono2002,Elzerman2004}. Simultaneous single-charge detection then reveals what the spin state was before the measurement. Real-time single-charge detection can be accomplished in several ways. In the first method, the conductance through a nearby charge detector is probed, either at baseband \cite{Vandersypen2004} or via RF modulation \cite{Barthel2009}. The charge detector can be a narrow channel called quantum point contact (QPC) or a small island that itself is capacitively coupled to the quantum dot or donor. In either case, the conductance through it directly depends on the charge occupation of the dot or donor (see Fig. 1 a,b). Alternatively, the ability of charges to move back and forth in response to an oscillating excitation can be probed. This amounts to an electrical susceptibility measurement, which is commonly implemented by looking at the reflection of an RF signal applied to one of the quantum dot gates \cite{Colles2013} or reservoirs \cite{ Gonzalez2015}. Single-shot measurement times down to 200 ns have been achieved in specific settings \cite{Barthel2010}, and read-out fidelities as high as 99.8\% have been reported \cite{Watson2015}.

Qubit reset or initialisation could be achieved by thermalisation to the ground state, but that would be very slow given that spin relaxation times are often in the millisecond to second range \cite{Hanson2007,Zwanenburg2013}. Faster approaches include initialization by measurement \cite{Nielsen2000} and spin-selective tunnelling from an electron reservoir or dot to a dot or donor \cite{Elzerman2004, Morello2010, Shulman2014}.

Finally, we note that microscopic variations in the semiconductor substrate and non-uniformities in the gate patterns lead to substantial variations from site to site in a realistic device. While progress has been made and high-quality double quantum dots have been reported \cite{Borselli2015}, an attractive but challenging solution would be to reach a uniformity level where a common (set of) DC voltage(s) would suffice to place each of several quantum dots in the desired configuration; e.g. systematically having a dot-to-dot variation in required gate voltage for single electron occupancy smaller than the charging energy. Donor fabrication introduces more challenges, but the strong confining potential can have specific advantages here due to the intrinsic large energy scales. Fabrication based on scanning-tunneling-microscopy (STM) \cite{Fuhrer2009} as compared to ion implantation has the further advantage that uncertainties in donor placement and capacitive coupling to nearby stray donors are significantly reduced. However, a systematic study on the relevant variations for a large array is missing. Furthermore, nominally identical operations currently require DC gate voltages, gate voltage pulses and microwave control signals that all differ in amplitude or duration from qubit to qubit. 

\section{Control signal requirements} 
\label{sec:reqs}

The discussion of electron spin qubits in quantum dots or donors leads us to the following commonly recurring requirements for the control signals. As can be seen from Fig.~\ref{fig:encodings}, not all requirements apply to each of the encodings, and this can be a criterion for comparing the merits of different encodings with each other. 

\begin{enumerate}
\item an independently calibrated and tuned DC gate voltage on every site (typically up to $\pm1$V)
\item independently calibrated and tuned gate voltage pulses on every site (typically up to tens of mV and with sub-ns rise times)
\item independently calibrated and tuned microwave magnetic or electric fields at every site (typically -40 to -20 dBm, 1-50 GHz bursts of 10 ns to 1 $\mu$s duration)
\item a high precision of each of the control signals to achieve error rates comfortably below the $1 \%$ accuracy threshold 
\item initialization, operations and read-out on timescales short compared to the relevant decoherence time.
\end{enumerate}

We now examine some of these requirements in more detail, and in particular consider which requirements can be relaxed. In the next section, we will present some general guidance for meeting the necessary requirements.

For the pulsed control signals, often only one of the two pulse stages requires precise tuning. For instance, the precise strength of the exchange interaction is important when the exchange is turned ``on", but the exchange strength in the ``off" state merely needs to be below some threshold, which is a much more relaxed constraint. Similarly, accurate level alignment is needed during read-out of a single spin based on spin-selective tunnelling to a reservoir \cite{Elzerman2004, Morello2010}, but when not reading out it suffices to stay in the regime with one electron per site. Spin read-out of two-electron spin states is typically even more forgiving, as it suffices to pulse from somewhere deep in the regime with one electron on each dot, to somewhere in the so-called pulse triangle with two electrons on one of the dots \cite{Petta2005, Prance2012}. Therefore, one could imagine that voltage pulses to, say, control exchange gates or initiate read-out can be made uniform across multiple (all) dots, by fine-tuning the exact qubit operating points via DC bias voltages. The main assumption in these examples is that the qubit is not sensitive to the exact DC gate voltage while in the ``off" state. As the qubit transition frequency may in fact vary with DC gate voltage \cite{Veldhorst2014, Veldhorst20152, Laucht2015}, unintentional single-qubit $\hat{z}$-rotations could occur and these must be tracked or corrected separately for every qubit.

For microwave control signals, we need to separately consider the microwave frequency versus amplitude and duration. The simplest approach is to assume that all qubits will need to be resonant with either a single frequency or a small number of frequencies. This can be achieved by $g$-factor control or Stark shifting, through either DC or pulsed control voltages \cite{Veldhorst2014, Laucht2015}, to bring qubits on specific sites in or out of resonance with the excitation. For conventional electron spin resonance (ESR) whereby a global microwave magnetic field is applied \cite{Koppens2006, Veldhorst2014, Pla2012, Muhonen2014}, the same microwave can be used to achieve the same angle of rotation on multiple qubits provided the amplitude variations are sufficiently small and the resonance frequency of all qubits resonant with the excitation is sufficiently uniform. Uncontrollable spin-orbit coupling renormalizing the $g$-factor can give qubit-to-qubit variations in the resonance frequency of order 10 to 100 MHz at $B$ = 1.5 T \cite{Veldhorst20152}. A possible strategy to overcome such variations is operating at significantly lower magnetic field.  Globally applied ac magnetics fields could give rise to excessive dissipation and heating, and the magnetic field profile may suffer from distortions due to all the metal interconnects. A strategy could be to integrate local microwave lines that are close to the qubits and only address subsections of the larger qubit array. Superconducting lines could further reduce dissipation.

For electric-dipole spin resonance, whether based on intrinsic spin-orbit interaction\cite{Nowack2007, Maurand2016} or on local magnetic field gradients to allow electric fields to drive spin transitions \cite{Tokura2006, Kawakami2014, Takeda2016}, dot-to-dot variations in the confining potential may impose different microwave amplitudes for every qubit. All-electrical control is often argued to be beneficial because of fast and local control. Essential in the design will be the interconnection between the microwave source and the individual qubits. Power dissipation will be significantly reduced compared to ESR, but avoiding cross-talk will be challenging. A solution for cross-talk could include to spatially separate qubits with equal resonance frequency.

The main message from this technical discussion is that even though some requirements can be relaxed, especially if the quantum dot properties are homogeneous, at least a subset of signals (DC, pulsed, or microwave) will need to be independently calibrated and delivered to each and every qubit. 

\section{Control signal wiring solutions}

How can we route qubit-specific classical control signals to a large number of quantum dot or donor qubits? The common understanding in the field is that directly connecting via wires or coax lines say $10^8$ sub-100 mK qubits to room temperature voltage sources, pulse sources and microwave sources, is impractical for several reasons. At the qubit chip level, it conflicts with Rent's rule in classical systems \cite{Lanzerotti2005} and practical limits to the number of pins on a chip. At the level of the transmission lines from room temperature to the chip, heat transport causes a heat load of a few mW on the 4 K plate. For comparison, cooling powers of currently used pulse tube systems are in the range of a few W at 4K. Below 4 K, superconducting lines can be used, which are poor thermal conductors and thus minimize heat load, but again power dissipated in the attenuators can heat up the coldest parts of the dilution refrigerator. A common view is that instead a combination of two ingredients will be required:
\begin{enumerate}
\item Multiplexing strategies
\item A first layer of classical electronics residing next to the qubits and commensurate with the inter-qubit spacing
\end{enumerate}
Other layers of classical electronics may reside farther away from the qubit plane and at higher temperature, as the data rates between layers higher up in the quantum computer architecture are orders of magnitude smaller than those between the physical qubits and the first control layer.

Within this framework, important choices include

\begin{enumerate}
\item What qubit density to work with? 
\item At what temperature do the qubits reside? Is operation at 1.5 K or 4 K possible? 
\item What is the functionality of the first electronics layer? 
\item What specifications must the electronics meet (clock speed, noise, resolution, frequency range, memory, power dissipation)?
\end{enumerate}

These questions are interrelated, for instance the qubit density and the cooling power (which depends on the temperature) impact the functionality and specifications of the electronics that can be achieved.
We next discuss platforms based on a dense qubit array or a sparse qubit array, and an operation temperature ranging from 100 mK to 4K.

\subsection{Dense qubit array and cross-bar addressing}

\begin{figure}
\includegraphics[width=0.3\textwidth]{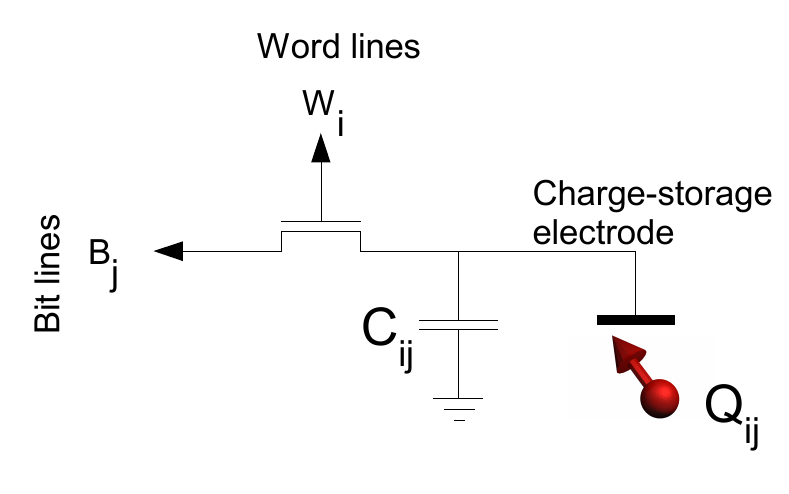}%
\caption{Charge-storage capacitors for biasing quantum dots, in analogy to DRAM.  Individual qubit communication can be achieved via a pair of word lines and bit lines. A voltage can be applied to qubit gate $Q_{ij}$ via $B_j$ by setting $W_i$ high and stored on capacitance $C_{ij}$ by subsequently setting $W_i$ low. Depending on the pitch and dimensions of transistors and quantum dots, more complex circuits can be constructed based on this method. 
\label{fig:1}}
\end{figure}

The most widely used mechanism for two-qubit gates using quantum dots is based on the exchange interaction \cite{Nowack2011, Brunner2011, Veldhorst2015}. This interaction couples the spin states of two electrons when their respective wave functions overlap, i.e. when the respective dots are tunnel coupled \cite{Loss1998}. The two-qubit exchange gate is very fast: it can be operated on sub-ns timescales, limited in practice by the bandwidth of the control electronics rather than by the underlying physics. In the absence of nuclear spin noise that is mostly relevant in III-V quantum dots \cite{Hanson2007}, the fidelity is often limited by electrical noise, usually charge noise from the amorphous materials and interfaces, and electrical noise on the gates.

Coherent spin exchange between neighbouring spins has been realised in double dots as well as in linear arrays of three dots \cite{Medford2013,Eng2015,Baart2016b}. Scaling a linear array up is relatively straightforward, but scaling to a large two-dimensional array of tunnel coupled dots is not. In order to have sufficient tunnel coupling between neighbours in the array, the center-to-center distance between dots must be no more than a few 100 nm in GaAs, less than 100 nm for qubits defined in silicon (a result of the the 5 times larger effective mass in Si), and less than 20 nm for donors (due to the strong confinement potential). Even taking a 100 nm qubit-to-qubit pitch as an example, a 2D array of 1 mm$^2$ would allow space for a massive $10^8$ qubits. However, at these densities and with at least one control line per qubit, fan-out requires multiple layers of interconnects. If the interconnects and control lines can come from all sides, but have the same width as the quantum dots, a $4 \times 4$ array requires two layers (the first layer to contact the outer dots, the second to contact the inner dots), five layers are needed for a $10 \times 10$ array and 50 layers for a $100 \times 100$ array. There is clearly a practical limit to the size of a monolithic array that can be wired up in this way. 

A possible strategy to partly overcome this fan-out problem is to borrow concepts from dynamic random access memory (DRAM). Rather than connecting every gate continuously to a voltage source, an individual gate is connected to a capacitor that stores the desired voltage. The voltages can be set efficiently via a cross-bar addressing scheme \cite{Veldhorst2016} (Fig. 3). Given that a single electron charge $e$ is the smallest amount of charge that can be added to the capacitor, the capacitance required to achieve a gate voltage resolution $\Delta V$ must be $C>|e|/\Delta V$. For $\Delta V = 1 \mu$V, this gives $C>160$ fF. Furthermore, thermal noise in the circuit when the switch is closed translates to an uncertainty in the gate voltage given by $V_{rms}=\sqrt{k_B T / C}$, which is a function of the capacitance but independent of the circuit resistance. Reaching a noise level $V_{rms} = 1 \mu$V would require temperatures below 10mK, or capacitances larger than 160 fF.

These charge-storage electrodes may have to be periodically refreshed, due to leakage or variations in the capacitive coupling to nearby structures. Such refreshing is routinely done in classical electronics. For instance, a typical refreshing interval time of DRAM is 64 ms where a refresh cycle is performed within 30 ns. If the 1\% weakest electrodes can be excluded, the interval time can be extended to a second. While the tolerances of quantum dot voltages are much more stringent, leakage is strongly reduced at a few Kelvin or below, so such an approach might be feasible. Experimental drifts of approximately one Coulomb oscillation per hour $\approx$ 8 mV/h have already been observed in charge-storage electrodes integrated with quantum devices \cite{Puddy2015}. However, more research is needed to demonstrate these drifts using electrodes that have a size comparable to the quantum dots and to minimize possible leakage pathways.

Globally controlling these floating electrodes could be done via an efficient cross-bar addressing scheme, using horizontal and vertical control lines that each have a spacing corresponding to the dot-to-dot distance. Assuming a dot-to-dot pitch of 50 nm, consistent with requirements for quantum dots, would imply an interconnect pitch of 50 nm, which is similar to what is possible with 14 nm node technology, the most advanced that is commercially available today \cite{Intel14nm}. Furthermore, 50 nm is below the 70 nm transistor gate pitch for the 14 nm node.  Therefore, unless dot dimensions can be kept slightly larger, integrating a single transistor above every quantum dot requires continued scaling of conventional CMOS devices, dictated by Moore's Law. 

A cross-bar approach can also provide a relatively economical avenue for qubit control. For instance, we can apply a voltage pulse on one of the vertical lines (combined with the DC voltage required by that site via a bias-tee) and use the horizontal line to select to which qubit the pulse is applied. As discussed in the section of control signal requirements, it should be possible to allow the same pulse amplitude to induce an exchange gate or initiate read-out across multiple dots. In this case, parallel addressing of multiple dots will be possible, as well as addressing for instance all dots or half of the dots (any combination of dots compatible with cross-bar selectivity is possible). It has indeed been shown that the cross-bar approach can be used to run the surface code, both in donor and dot platforms \cite{Hill2015, Veldhorst2016}. It was also shown that surface code variations can be implemented with reduced local control \cite{Pica2016, Gorman2016}.

Initiating parallel read-out is possible with a cross-bar approach as well, with vertical lines used to select the set of qubits underneath and horizontal lines used to carry the corresponding read-out signals. It may be possible to re-use the same cross-bar that is used for control, also for read-out, for instance using dispersive gate read-out \cite{Colles2013}. An RF signal is then applied to a vertical line (again added to a DC gate voltage) and the horizontal lines select the qubit that is read out. This procedure comes at a cost. In its simplest form, an array of $N$ qubits requires $\sqrt{N}$ repetitions of this read-out protocol to measure all the qubits.  

This slow-down has two sides. First, it requires that probability of error of a qubit during $\sqrt{N}$ read-out cycles stays far below the accuracy threshold. Here the extremely long memory times of spin qubits under dynamical decoupling, of order one second \cite{Veldhorst2014, Muhonen2014}, are crucial. Second, it slows down the net clock cycle of the surface code operation by a factor $\sqrt{N}$. Here we note that it is not clear what the optimal effective clock cycle is. Too slow is not good since it slows down the computation. Too fast is not good either, since then the classical processors cannot keep up processing the massive data streams produced by the surface code syndrome measurements, and this will pose a hard boundary. This flexibility in choosing the clock cycle of the classical computer may turn out to be an important advantage of electron spin qubits over e.g. superconducting qubits.

As a final note, using RF techniques for read-out \cite{Colles2013, Gonzalez2015}, it may be possible to combine the cross-bar approach with frequency multiplexing \cite{Hornibrook2014}, so that each horizontal line can carry multiple read-out signals simultaneously. The demonstrated on-chip resonators \cite{Hornibrook2014} will be challenging to fit locally into a dense array. However, frequency multiplexing could also be achieved by clever crossbar operation. For example, if $J$-gates are connected to vertical lines, these can be frequency-modulated so that each vertical line has a different modulation frequency. The resonance frequency of the readout circuits, measured along the horizontal lines, will then shift corresponding to the respective modulation frequency. This frequency multiplexing enables simultaneous read-out along horizontal lines. Global simultaneous read-out is then obtained by connecting each horizontal line to a separate circuit or by frequency multiplexing each horizontal line. If $k$ frequencies can be simultaneously read out, $k \sqrt{N}$ qubits can be read out in parallel. This gives further design flexibility and room for optimisation.

\begin{figure*}
\includegraphics[width=0.9\textwidth]{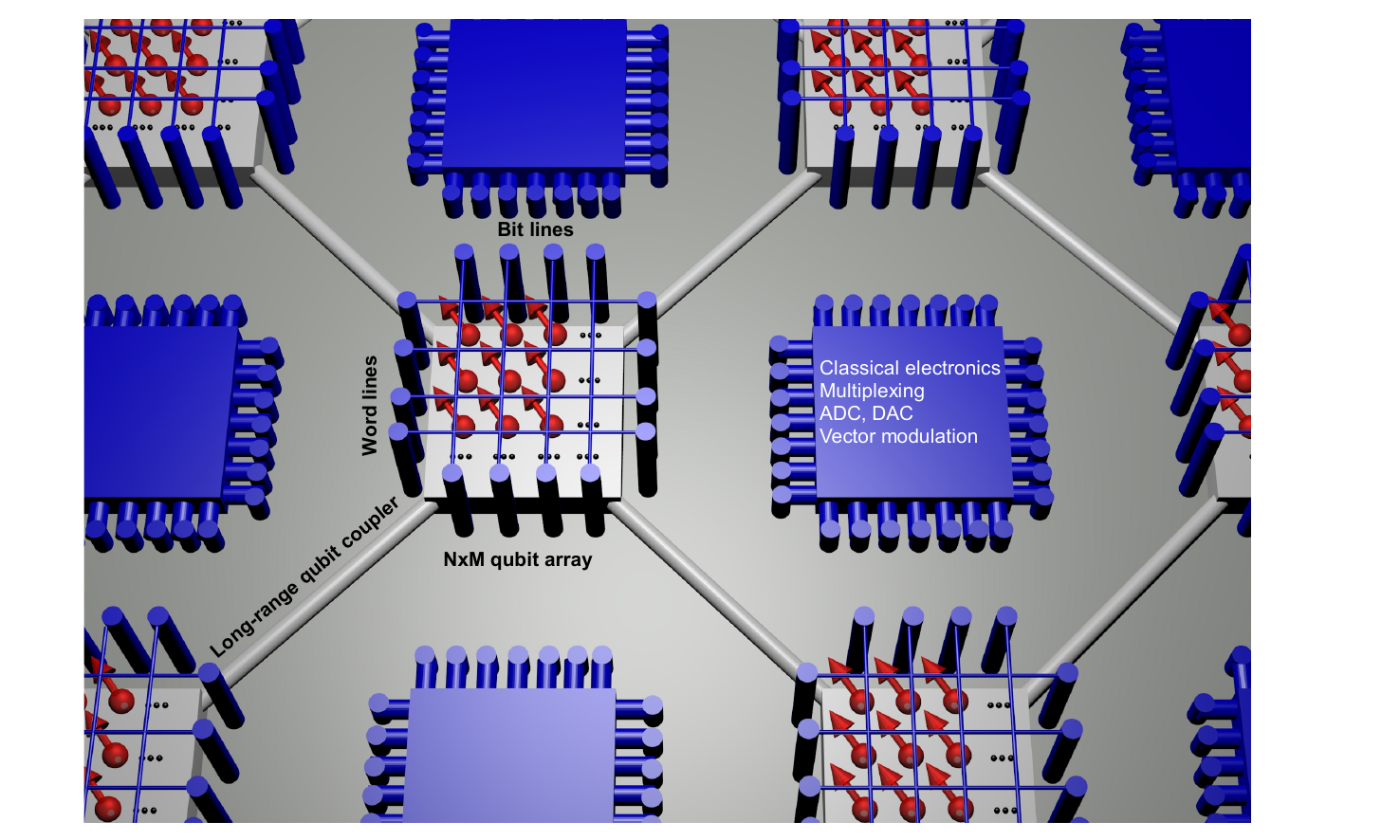} 
\caption{Sparse qubit array with local electronics. Long-distance qubit coupling opens up space for local electronics that can control a small dense qubit array. In the schematic, this electronics is placed in the qubit plane. Alternatively, it could be located on a separate chip and connected to the qubit chip by flip-chip or similar technologies. A crucial aspect is the optimum qubit array size $N \times M$ and the functionality of the local electronics. Ideally, the local electronics include ADC and DAC converters, as well as vector modulation, such that a minimal number of control lines needs to interface with the outside. Giving the strong dependence of refrigerator cooling power on temperature, power dissipation in the classical electronics integrated with the qubits would likely require the qubits to operate at higher temperatures. Therefore, the demonstration of high-fidelity spin qubit operation at four Kelvin would be a milestone towards extendable structures. 
\label{fig:2}}
\end{figure*}

\subsection{Sparse qubit arrays and local electronics}

Several alternative spin qubit coupling mechanisms exist besides direct exchange coupling, that allow the building of two-dimensional spin qubit arrays without the need for direct tunnel coupling between neighbouring qubits in four directions (north, south, east, west). Many of these mechanisms have in common that they allow the separation of the qubits by larger distances, varying from roughly one $\mu$m to roughly one mm. Proposals for coupling spin qubits at a distance rely on the use of superconducting resonators \cite{Burkard2006, Childress2004, Hu2012, Viennot2015}, capacitive coupling \cite{Shulman2012,Trifunovic2012,Tosi2016}, ferromagnets \cite{Trifunovic2013}, superconductors \cite{Leijnse2013, Hassler2015}, intermediate dots or dot arrays \cite{Friesen2007, Trauzettel2007, Stano2015, Srinivasa2015, Baart2016b}, or surface acoustic wave cavities \cite{Schuetz2015}. An alternative approach consists of shuttling electrons across the chip between distant quantum dots, where the electrons are propelled by time-varying gate voltages \cite{Taylor2005a, Baart2016a} or surface acoustic waves \cite{Hermelin2011, McNeil2011}.

Widely spaced qubit arrays can alleviate fan-out and wiring problems, simply by allowing more space for routing as shown in Fig. 4. Yet, even if this allows space to connect each qubit to one or more control lines running off the chip, we mentioned before that connecting individual qubits to sources and generators a large distance away is not viable. Therefore, the more important advantage of space between the qubits may be that it allows a first layer of control electronics that is commensurate with the inter-qubit spacing to be placed directly above or in the qubit layer. If placed above the qubit layer, this classical layer can be interfaced with the qubit layer via an interposer, flip-chip (C4) technology or similar methods. Thermal isolation between the quantum and classical chips could be provided by using superconducting vias for connection. In this way, heating of the qubits by thermal dissipation in the classical circuitry is minimized.
When transistors are realized in the same plane as the qubit layer, they could be integrated directly with traditional CMOS fabrication.

Depending on the actual spacing between qubit arrays and on the power budget, the functionality of the classical layer can be more or less advanced \cite{Hornibrook2015}. At the lowest level, simple multiplexing strategies based on switches can be implemented. What would have more impact is if ADC, DAC, and vector modulation could be implemented locally in the first classical layer. In this case, only digital signals must flow between the first classical layer and a second layer higher up in the control structure, potentially even at room temperature,  where the digital data is processed. The required bandwidth of the communication channel between the classical layers is then much smaller, as per qubit one or a few bits of information must be transmitted per clock cycle, instead of time traces containing a large number of analog data points. Even then, data rates to room temperature are substantial. For example, $10^8$ qubits at 2 MHz clock speed gives 100 Tb/s data rates, if every second each qubit is read out and provides one bit of information. Control will require a few bits and several operations per surface code cycle. Therefore, local error decoding would be highly attractive but also most demanding in terms of circuit complexity.

Several open questions will determine the feasibility of this approach. First, how many transistors are needed for each functionality and how many classical clock cycles per quantum clock cycle can we afford to avoid a communication backlog? Second, what is the required circuit complexity to implement (part of) the error decoding logic locally? Third, how low can dissipation per transistor be in optimised, low-power circuits operating at cryogenic temperatures? Fourth, how large can we make the cooling power of future refrigerators at various temperatures? Fifth, what are convenient qubit spacings to allow reliable gate operations from the physics perspective? Sixth, up to what temperature can spin qubits be operated without excessive compromises in the fidelity of initialisation, coherent operations, memory time, and read-out? We explore this last question in more detail in the next section.

\subsection{Hot qubits}
Much would be gained by qubits that can operate at 1 K to 4 K. At 4 K, the cooling power of a single commercial pulse tube cooler as used in qubit experiments today is 1-2 W. By comparison, powerful dilution refrigerators offer a cooling power of 1 mW at 100 mK. At $T<100$ mK, we therefore expect that only very simple functionality can be realised without excessive heat dissipation. Superconducting classical circuits \cite{Likharev1991} dissipate very little power, but are complex in design, lacks the memory function, and have a large footprint. Operating spin qubits at 4 K, with a thousand-fold increase in available cooling power, makes the prospect of electronics commensurate with and right next to the qubit plane more realistic. An integrated quantum-classical structure would have multiple advantages in solving the fan-out problem, would simplify the RF wiring and reduce signal losses. 

A major attraction of Si-MOS based quantum dots and donor-based qubits is that they can have energy scales that are compatible with 1 K to 4 K operation. Proper operation requires that the relevant energy scales are about five times larger than the thermal energy, which is 340 $\mu$eV at 4 K. Charging energies of donors and small quantum dots are easily in excess of 10 meV and orbital energies can be of order 10 meV as well\cite{YangPRB2012}, satisfying this requirement. However, in silicon there is also a valley degree of freedom. Silicon has a six-fold degeneracy due to crystal symmetry, which is broken at the interface leaving two relevant valley states. These lowest-energy valley states can be split via a sharp confinement potential, e.g. the silicon-SiO$_2$ or Si/SiGe interface, and a vertical electric field. In Si-MOS dots, the valley splitting has reached almost 1 meV \cite{Yang2013, Veldhorst2014}, and could be pushed up further by reducing the device dimensions and increasing the electric field by confinement gates. This would allow initialization in the lowest-energy orbital and valley state. Conventional single-spin read-out requires a Zeeman splitting several times larger than the thermal energy \cite{Elzerman2004}. This approach would not be viable at these temperatures, as simply increasing the magnetic field would imply impracticable qubit operation frequencies of (sub) THz and potentially too short relaxation times.

High-fidelity initialization and read-out of spin states can make use of the single-dot singlet-triplet splitting, which is typically somewhat below the valley splitting due to the exchange interaction \cite{Hanson2007}. In this scenario, two electrons are loaded on the same dot, occupying the ground state valley and orbital state with the spins in a spin singlet configuration. One electron is then moved to the neighbouring dot by adjusting the gate voltages, creating a state with one electron on each dot. If the movement is diabatic with respect to the difference in Zeeman energy between the dots, the spins will remain in their spin state and thus be initialized in the singlet state, which is a natural initial state for a $S-T_0$ qubit\cite{Petta2005, Maune2012}. When using $\{\ket{\uparrow},\ket{\downarrow}\}$ qubits, the spin singlet can be rotated to $\ket{\uparrow}\ket{\uparrow}$ if needed. If the difference in Zeeman energy is large compared to the exchange energy, diabatic pulsing might not be an option. Instead, adiabatic transfer of one electron to the neighbouring dot will result in the $\ket{\uparrow}\ket{\downarrow}$ state \cite{Petta2005} (Fig. \ref{fig:encodings}). 

With one well-initialized electron on each dot, qubit splittings can be chosen in a comfortable range, say 5-200 $\mu$eV, which corresponds to accessible microwave frequencies of 1-50 GHz. Hence by combining a large energy splitting for initialization and read-out with a lower level splitting during qubit manipulation, the frequencies for driving qubits do not have to be scaled up with the operating temperature.

The spin relaxation time $T_1$ will be reduced with higher temperature. Below 100 mK, $T_1$ is typically very long, especially in silicon, with measured $T_1$ times of over one second \cite{Morello2010, Yang2013}; see \cite{Tahan2014} for a theoretical analysis on the limiting relaxation mechanisms. At low temperature, the temperature dependence of $T_1$ is dictated by one-phonon (direct) processes, and the relaxation rate will increase roughly linearly with temperature \cite{Hanson2007}. However, the relaxation rate can have a much stronger temperature dependence at higher temperatures due to two-phonon transitions, such as $1/T_1 \propto T^{7-9}$ (Raman) and/or $1/T_1 \propto e^{-\Delta E/k_B T}$ (Orbach), where $\Delta E$ is the energy to the first orbital state. For donors, the transition to the exponential temperature dependence due to Orbach transitions occurs at 6 K for phosphorus, 11 K for arsenic, 4 K for antimony, and 26 K for bismuth, all at a magnetic field of 0.3 T. The measured $T_1$ is above one second at 4 K in all cases \cite{Castner1962}. For silicon quantum dots, there are few experimental reports on the temperature dependence of $T_1$ \cite{Shankar2010}. Based on the large orbital splitting of order 10 meV that can be realized in silicon quantum dots \cite{YangPRB2012}, one would expect the transitions to two-phonon processes to occur at relatively high temperatures as well. However, imperfect interfaces give rise to spin-orbit coupling between the valley states, and this opens a new channel for relaxation as observed in experiment \cite{Yang2013}, which will have a strong sample-to-sample dependence. Nevertheless, long $T_1$ times have been achieved even in systems with very small valley splitting \cite{Kawakami2014}. This suggests that at least in this temperature range, multi-phonon processes do not dominate and more research on the temperature dependence is needed. Nonetheless, the long relaxation times leave a lot of margin, and we anticipate that it is possible to substantially increase the operating temperature of silicon spin qubits.

Decoherence from hyperfine interaction with nuclear spins in the substrate will be approximately temperature independent. An important question is to what extent charge noise will be enhanced by thermal excitations. Established models indicate that the noise increases linearly with T, and such signatures are seen in recent experiments on SiGe and SiMOS dots \cite{Freeman2016}. Charge noise affects spin states most strongly during gates based on exchange or capacitive coupling (Fig. 2), but also a single spin is sensitive to electric fields through the Stark effect, and this sensitivity is higher if local magnetic field gradients are present. Significant improvements in the quality of exchange oscillations (the basis for most two-qubit gates, and for single-qubit gates in some qubit representations) were recently obtained by keeping the qubits at all times at the so-called symmetry point (Fig. \ref{fig:encodings}a) \cite{Reed2016,Martins2016}. At this operating point, the spin states are to first order insensitive to the energy detuning between neighbouring dots. This detuning is typically the main channel through which charge noise affects the qubit splitting. Even coupling spin qubits via resonators may be possible at 4 K, despite the fact that the resonator will be thermally populated\cite{Schuetz2016}. Altogether, we believe that potential 4 K operation of spin qubits is an attractive possibility.

\section{Conclusions }

Wiring up large qubit arrays is a common, central challenge across all qubit platforms. From the above discussion, we see that electron spin qubits in quantum dots or donors offer several particularly attractive features for overcoming this challenge. First, the sub-100 nm lateral dimensions of quantum dots or donors allow for highly dense qubit registers that nevertheless can be wired up with multiplexing and cross-bar approaches with charge-storage electrodes. The feasibility of such approaches strongly benefits from the extremely long coherence times of electron spins in nuclear-spin-free host materials such as isotopically purified $^{28}$Si \cite{Tyryshkin2012, Veldhorst2014, Muhonen2014} which relax the requirements of parallel read-out and control that short-lived qubits must meet. Second, multiple ideas have been proposed for interconnecting qubit arrays over micron to mm distances. This leaves flexible space for interconnects and integrated electronics. Third, spin qubits on dots or donors may be operated at temperatures of 1-4 Kelvin, where the available cooling power is about 1000 times larger than below 100 mK, the typical operating temperature today. This would greatly simplify the integration of a first layer of classical control electronics right next to the qubits, again strongly relaxing the interfacing challenges.

These proposed solutions and approaches are not mutually exclusive. For instance, charge-storage electrodes can be beneficial also in sparse arrays, and a classical layer with (very) limited functionality could be incorporated with dense arrays. Furthermore, it is clear that there is still a big step to take from formulating general ideas as done here, to a complete proposal for an actual device, including device lay-outs, dimensions, power budgets, and so forth. Nevertheless, it is clear that spin qubits offer several particularly attractive possibilities in this direction. Finally, the continuous development of semiconductor technology provides further perspective that the wiring challenges can in fact be overcome, paving the way for the construction of a large-scale universal quantum computer.

\end{document}